\documentclass[12pt]{article}

\usepackage{amsmath}
\usepackage{amsfonts}
\newcommand {\Pbar}{{\mbox{\rm$\mbox{I}\!\mbox{P}$}}}

\newsavebox{\zzzbar}
\sbox{\zzzbar}
  {\setlength{\unitlength}{0.9em}
  \begin{picture}(0.6,0.7)
  \thinlines
  \put(0,0){\line(1,0){0.6}}
  \put(0,0.75){\line(1,0){0.575}}
  \multiput(0,0)(0.0125,0.025){30}{\rule{0.3pt}{0.3pt}}
  \multiput(0.2,0)(0.0125,0.025){30}{\rule{0.3pt}{0.3pt}}
  \put(0,0.75){\line(0,-1){0.15}}
  \put(0.015,0.75){\line(0,-1){0.1}}
  \put(0.03,0.75){\line(0,-1){0.075}}
  \put(0.045,0.75){\line(0,-1){0.05}}
  \put(0.05,0.75){\line(0,-1){0.025}}
  \put(0.6,0){\line(0,1){0.15}}
  \put(0.585,0){\line(0,1){0.1}}
  \put(0.57,0){\line(0,1){0.075}}
  \put(0.555,0){\line(0,1){0.05}}
  \put(0.55,0){\line(0,1){0.025}}
  \end{picture}}

\newsavebox{\uuunit}
\sbox{\uuunit}
    {\setlength{\unitlength}{0.825em}
     \begin{picture}(0.6,0.7)
        \thinlines
        \put(0,0){\line(1,0){0.5}}
        \put(0.15,0){\line(0,1){0.7}}
        \put(0.35,0){\line(0,1){0.8}}
       \multiput(0.3,0.8)(-0.04,-0.02){12}{\rule{0.5pt}{0.5pt}}
     \end {picture}}

\newcommand{\QED}{{\hspace*{\fill}\rule{2mm}{2mm}\linebreak}}

\newtheorem{definition}{Definition}[section]
\newtheorem{proposition}[definition]{Proposition}
\newtheorem{theorem}[definition]{Theorem}
\newtheorem{lemma}[definition]{Lemma}
\newtheorem{corollary}[definition]{Corollary}

\newcommand{\Z}{\Bbb Z}
\newcommand{\R}{\Bbb R}
\newcommand{\N}{\Bbb N}

\newcommand{\pee}{\Pbar}

\newcommand{\re}{\ensuremath{\mathcal{R}}}
\newcommand{\ce}{\ensuremath{\mathcal{C}}}
\newcommand{\s}{\ensuremath{\mathcal{S} }}

\newcommand{\aaa}{\ensuremath{\mathcal{A}}}

\begin{document}
\pagestyle{myheadings} \markright{The abelian sandpile}
\setlength{\textheight}{21cm}
\title{{\bf The Abelian sandpile; a mathematical introduction}}
\author{Ronald Meester\footnote{ Faculty of Exact Sciences, Free University Amsterdam, de Boelelaan 1081,
1081 HV Amsterdam, The Netherlands},
Frank Redig\footnote{ Technische Universiteit Eindhoven, Faculteit Wiskunde en Informatica, Postbus 513, 5600 MB Eindhoven, The Netherlands} and Dmitri
Znamenski$^*$}
\date{June 12, 2001}
\maketitle

\begin{abstract}
We give a simple rigourous treatment of the classical results of
the abelian sandpile model. Although we treat results which are well-known in the physics literature, in many cases we did not find complete proofs in the literature.
The paper tries to fill the gap between the mathematics and the physics literature on this subject, and also presents some new proofs. It can also serve as an introduction to the model.
\\[5mm]
\end{abstract}
\begin{flushleft}
{\bf keywords:} abelian sandpile, recurrent configurations, burning algorithm.
\\[5mm]
\end{flushleft}
\vspace{12pt}

\section{Introduction}
Since its introduction in \cite{Bak}, the abelian sandpile model has been one of the archetype
models
of self-organized criticality. In words, the model can loosely be described as
follows. Each vertex in some finite subset $V$ of the $d$-dimensional integer lattice contains
a certain number of sand grains. At discrete times, we
add a sand grain to a randomly chosen vertex in $V$. Each vertex has a maximal capacity of
sand grains, and when we add a grain to a vertex which has already reached this maximal
capacity,
grains of this site move to the neighbouring vertices, starting an {\sl avalanche}. This
moving of grains to neighbours is called a toppling and it can in turn  cause neighbouring
vertices to exceed their capacity. In this case, these neighbouring vertices send their grains
to their neighbours, etcetera. At the boundary, grains are lost.
The avalanche continues as long as there is at least one vertex
which exceeds its capacity. A configuration in which no vertex exceeds it capacity is called
stable.

Physicists are interested in the statistics of avalanches,
see \cite{Dhararch}.  They study the
size and duration of these avalanches, and try to describe them in terms of power laws (see e.g.\ \cite{IP}). The spatial correlations in the stationary state are also believed to decay as a power law. For some
particular observables this has been proved see e.g. \cite{Dhararch},
and references therein. The presence of power law decay of correlations - typical for models at the critical point without ``fine tuning"
of parameters (such as temperature or magnetic field)- has led to the term ``self-organized criticality". This means that the dynamics, a combination
of external driving (adding grains) and relaxation, drives the system into
a state which resembles a statistical mechanical model at the critical point.
In a variety of natural phenomena (e.g. mountain heights, earthquakes) power law
decay of correlations is observed empirically. The BTW-model shows how a
simple driven dynamics can explain this behavior: the system is naturally
driven
into a state where no natural (finite) correlation length can be defined.

The abelian sandpile model allows, to some extent at least, for rigorous mathematical
analysis.
It can be described in terms of an abelian group of addition operators. The abelianness is an essential simplifying property, which allows for many exact results.
We noted, however, that many results in the physics literature that are claimed as being
exact, are not always rigorous and/or complete. Sometimes, it turns out that
the ideas can be turned into a rigorous proof simply by being a bit more precise. But
sometimes, it seems that more is needed to do that. Since we think it is important that
mathematicians take up the subject of self-organized criticality, we want to make sure that at
least in the basic model of self-organized criticality, there is a reference containing a
mathematically rigorous analysis of the model.
We hope and expect that this note increases the interest of mathematicians for self-organized
criticality.  We treat the following aspects.

First, we consider the abelianness of the model. It will be clear from the precise definition
of the model below, that if two vertices $x$ and $y$ exceed their capacity, and we only
topple these two vertices (so we do not topple vertices which exceed their capacity as
a result of the toppling of $x$ and/or $y$), then it doesn't matter in which order we do this:
the resulting configuration after toppling $x$ and $y$, and only these, is always the
same. This elementary fact does {\sl not} imply that if we have multiple vertices exceeding
their capacity, then the final {\sl stable} configuration, obtained by toppling until no
vertex exceeds its capacity anymore, is independent of the order in which we topple. Indeed, by toppling $x$ first,
say, we have to take into account the possibility that a certain vertex needs to be toppled,
which would never have been toppled, if $y$ had been toppled first. The essential point is to prove that irrespective of the order in which we
perform the topplings, {\sl the same sites are toppled the same number
of times}.

After having proved the abelian property,  we define the Markov chain
associated with the sandpile model. In Section 4, we investigate the recurrent configurations
of this Markov chain, and show that Dhar's definition of recurrence (see
\cite{Dhar3}) is in this case
the same as classical recurrence in the language of Markov chains. The number of recurrent configurations is proved to equal the number of
group elements of the ``group of addition operators". Our proof is in the spirit of \cite{Dhar3}.

Finally we deal with the relation between so called ``allowed" and recurrent configurations.
We shall call a configuration allowed if it passes a certain test via the well known {\sl burning algorithm}. The equivalence between allowed and recurrent was open in \cite{Dhar3}, and has been settled via a correspondence between allowed configurations and spanning trees in \cite{IP}. We give
an alternative proof of the equivalence allowed/recurrent, not using
spanning trees.

\section{The model}
Let $V$ be a finite subset of $\Z^d$. An integer valued matrix $\Delta^V_{x,y}$
indexed by the sites of $x,y\in V$ is a {\it toppling matrix} if it satisfies the following
conditions:
\begin{enumerate}
\item
For all $x,y\in V$, $x\not=y$, $\Delta_{x,y}^V=\Delta_{y,x}^V\leq 0$,
\item
For all $x\in V$, $\Delta_{x,x}^V\geq 1$,
\item
For all $x\in V$, $\sum_{y\in V} \Delta_{x,y}^V \geq 0$,
\item
$\sum_{x,y\in V} \Delta_{x,y}^V > 0.$
\end{enumerate}
The fourth
condition ensures that there are sites (so-called {\sl dissipative
sites}) for which the inequality in the third condition is strict. This
is fundamental for having a well defined toppling rule later on. In the
rest of the paper we will choose $\Delta^V$ to be the lattice Laplacian
with open boundary conditions. More explicitly:
\begin{eqnarray}\label{laplace}
\Delta_{x,x}^V&=& 2d \ \mbox{ if}\ x \in V,\nonumber\\
\Delta_{x,y}^V&=& -1 \ \mbox{ if}\ x\  \mbox{and}  \ y \ \mbox{are
nearest  neighbors,}\nonumber\\
\Delta_{x,y}^V &=& 0 \ \mbox{otherwise}.
\end{eqnarray}
The dissipative sites then correspond to the boundary sites of $V$. This restriction is for
convenience only: the essential features on which
proofs are based are symmetry and existence of dissipative sites.

\subsection{Configurations}
A {\sl height configuration} $\eta$ is a mapping from $V$ to
$\N=\{1,2,...\}$ assigning to each site a natural number $\eta (x) \geq 1$
(``the number of sand grains" at site $x$). A configuration $\eta \in
\N^V$ is called {\sl stable} if, for all $x\in V$, $\eta(x)\leq
\Delta_{x,x}^V$. Otherwise $\eta$ is {\sl unstable}. We denote by
$\Omega_V$ the set of all stable height configurations. The maximal element
of $\Omega_V$ is denoted by $\eta^{max}$ ( i.e., $\eta^{max} (x) = \Delta^V_{x,x}$ for all
$x\in V$ ). For  $\eta \in
\N^V$ and $V'\subset V$, $\eta|_{V'}$ denotes the restriction of $\eta$
to $V'$.

\subsection{The toppling rule}
The {\sl toppling rules } corresponding to the toppling matrix $\Delta^V$
are the mappings $T_x$
\[
T_x: \N^V\rightarrow \N^V,
\]
indexed by $V$, and defined by
\begin{eqnarray}\label{Tdelta}
T_x(\eta) (y) &=& \eta (y) -\Delta_{x,y}^V \ \mbox{ if}
\ \eta(x) > \Delta_{x,x}^V,\nonumber\\
&=& \eta (y) \ \mbox{ otherwise}.
\end{eqnarray}
In words, site $x$ topples if and only if its height is strictly larger
than $\Delta_{x,x}^V$, by transferring $-\Delta_{x,y}^V$ grains to site
$y\not= x$ and losing itself $\Delta_{x,x}^V$ grains. Toppling rules
commute on unstable configurations. This means for $x,z\in V$ and $\eta$
such that $\eta(x) > \Delta_{x,x}^V$ and $\eta(z) > \Delta_{z,z}^V$,
\begin{equation}\label{abrule}
T_x\circ T_z (\eta)=T_z\circ T_x(\eta).
\end{equation}
Choose some enumeration $\{ x_1,\ldots,x_n \}$ of the set $V$. The {\sl
toppling transformation} is the mapping
\[
\mathcal{T}_{\Delta^V}: \N^V\rightarrow \Omega_V
\]
defined by
\begin{equation}\label{deftop}
\mathcal{T}_{\Delta^V}(\eta ) = \prod_{i=1}^N T_{x_i}(\eta)
\end{equation}
{\bf Remark:}\\
It is not clear that the requirement $\mathcal{T}_{\Delta^V}(\eta)\in\Omega_V$
together with (\ref{deftop}) defines the toppling transformation uniquely.
The first problem could be that $N$ in (\ref{deftop}) is not finite. By
the presence of dissipative sites this cannot happen, i.e., for any unstable
configuration $\eta$ there exists $(x_1,\ldots,x_N)$ such that
$\prod_{i=1}^N T_{x_i}(\eta)$ is stable.
The second problem is whether the $N$-tuple $(x_1,\ldots,x_N\in V^N$ is unique
up to permutations. It is precisely the content of the next section to prove this
fact.

\subsection{The abelian property}\label{abel}
In this section, we shall prove that equation (\ref{deftop})  properly defines
a transformation from unstable to stable configurations.

\begin{theorem}
The operator ${\cal T}_{\Delta^V}$ is well defined.
\end{theorem}
{\bf Proof:}
Suppose that a certain configuration $\eta$ has more than one unstable site.
In that situation, the order of the topplings is not fixed. Clearly,
if we only topple site $x$ and site $y$, the order of these
two topplings doesn't matter and both orders yield the same result. In the physics literature,
this is often presented as a proof that ${\cal T}_{\Delta^V}$  is well defined. But clearly,
more is needed to guarantee this. The problem
is that toppling $x$ first, say, could possibly lead to a new
unstable site $z$, which would never have become unstable if $y$ had
been toppled first. This is the key problem we have to address. More precisely, we  have to
prove
the following statement: no matter in which order we perform topplings, we always topple the
same sites the same number of times, and thus obtain the same final configuration. Our proof is
inductive, and runs as follows.

Let $\eta$ be an unstable configuration, and suppose that
$$
T_{x_N}\circ \cdots \circ T_{x_2}\circ T_{x_1}(\eta)
$$
and
$$
T_{y_M}\circ \cdots \circ T_{y_2}\circ T_{y_1}(\eta)
$$
are both stable, and both sequences are minimal in the sense that
$
T_{x_i}\circ \cdots \circ T_{x_2}\circ T_{x_1}(\eta)
$
and
$
T_{y_j} \circ \cdots \circ T_{y_2}\circ T_{y_1}(\eta)
$
are not stable, for all $i < N$ and $j < M$. We need to show that $M=N$, and that the sequences
$x_1,x_2,\ldots,x_N$ and $y_1,y_2,\ldots,y_N$ are permutations of each other.
To do this, we choose $N$ minimal with the property that there exists a sequence $x_1, \ldots, x_N$
with the property that $T_{x_N}\circ \cdots \circ  T_{x_2}\circ T_{x_1}(\eta)$ is stable. We now perform induction with respect to $N$. For $N=1$, there is nothing to prove. Suppose now that $N>1$ and
that the result has been shown for minimal length $N-1$.
Let $y_1,y_2,\ldots,y_M$ be a sequence so that
 $T_{y_M}\circ \cdots \circ T_{y_2}\circ T_{y_1}(\eta)$ is stable.
Since $\eta(x_1)>\Delta_{x_1,x_1}$, $x_1$ must appear at least once in the sequence $y_1,y_2,\ldots,y_M$. Choose  $k$ minimal so that $y_k=x_1$. Now we claim that
$$T_{y_M}\circ \cdots \circ T_{y_{k+1}}\circ T_{x_1}\circ T_{y_{k-1}}\circ \cdots \circ T_{y_2}\circ T_{y_1}(\eta)$$
and
$$T_{y_M}\circ \cdots \circ T_{y_{k+1}}\circ T_{y_{k-1}}\circ T_{x_1}\circ \cdots \circ T_{y_2}\circ T_{y_1}(\eta)$$
are the same.
To see this, define $\eta'=T_{y_{k-2}}\circ \cdots \circ T_{y_2}\circ T_{y_1}(\eta)$. $x_1$ has not been toppled at this point, hence $\eta'(x_1)>\Delta_{x_1,x_1}$. We also have  $\eta'(y_{k-1})>
\Delta_{y_{k-1},y_{k-1}}$,  and therefore we are allowed to interchange $T_{x_1}$ and
$T_{y_{k-1}}$. Repeating this argument, we can transfer $T_{x_1}$ to the right completely, and this leads to the conclusion that
$$
T_{y_M}\circ \cdots \circ T_{y_{k+1}}\circ T_{y_k}\circ T_{y_{k-1}}\circ \cdots\circ T_{y_2}\circ T_{y_1}(\eta)
$$
and
$$
T_{y_M}\circ \cdots \circ T_{y_{k+1}}\circ T_{y_{k-1}}\circ \cdots \circ T_{y_1}\circ T_{x_1}(\eta)
$$
are the same stable configuration. Now apply the induction hypothesis to $T_{x_1}(\eta)$
and the proof is complete.
\QED

\subsection{Addition operators}
For $\eta\in \N^V$ and $x\in V$, let $\eta^x$ denote the configuration
obtained from $\eta$ by adding one grain to site $x$, i.e. $\eta^x
(y)=\eta (y) +\delta_{x,y}$. The {\sl addition operator} defined by
\begin{equation}\label{axV}
a_{x,V}:\Omega_V\rightarrow\Omega_V;\eta\mapsto a_{x,V}\eta =
\mathcal{T}_{\Delta^V} (\eta^x)
\end{equation}
represents the effect of adding a grain to the stable configuration
$\eta$ and letting the system topple until a new stable configuration is
obtained. By abelianness, the composition of addition operators is
commutative: for all $\eta \in \Omega_V,\,x,y\in V$,
\[
a_{x,V} (a_{y,V}\eta ) = a_{y,V} (a_{x,V} \eta ).
\]

\subsection{The Markov chain}
Let $p$ denote a probability measure on $V$ with support $V$, i.e. numbers
$p_x$, $0< p_x< 1$ with $\sum_{x\in V} p_x =1$. We define a discrete time
Markov chain $\{\eta_n: n\geq 0\}$ on $\Omega_V$ by picking a point $x\in
V$ according to $p$ at each discrete time step and applying the addition
operator $a_{x,V}$ to the configuration. This Markov chain has the
transition operator
\begin{equation}\label{2.9}
P_V f(\eta ) = \sum_{x\in V}p_x f(a_{x,V}\eta ).
\end{equation}
We will denote by $\pee_\eta$ the Markov measure of the chain with transition
operator $P_V$ starting from $\eta$.

A configuration $\eta\in\Omega_V$ is called {\sl recurrent} for the (discrete) Markov
chain if
\begin{equation}\label{recdef}
\pee_\eta \left( \eta_n = \eta \ \mbox{ for infinitely many} \ n\right) = 1.
\end{equation}
A configuration which is not recurrent is called {\sl transient}.
Let us denote by $\re_V$ the set of all recurrent configurations of the Markov chain with
transition operator (\ref{2.9}). As we will show later on, this set is independent of the
chosen $p_x$, as long as $p_x>0$ for all $x$.

Let $\eta,\zeta\in\Omega_V$. We say that $\zeta$ can be reached from $\eta$
in the Markov chain (notation $\eta\hookrightarrow\zeta$) if there exists
$n\in \N$ such that $\pee_\eta (\eta_n = \zeta ) >0$.
Two configurations $\eta,\zeta\in\Omega_V$ are said to communicate in the
Markov chain (notation $\eta\sim\zeta$) if $\eta\hookrightarrow\zeta$
and $\zeta\hookrightarrow\eta$.
The relation $\sim$ defines an equivalence relation on configurations, which satisfies the
following property:
if $\eta\in\re_V$ and $\eta\sim\zeta$, then $\zeta\in\re_V$. In fact, every
configuration that can be reached from a recurrent configuration is recurrent, and hence on
$\re_V$ the relations $\hookrightarrow$ and $\sim$
coincide. The set $\re_V$ can be partitioned into equivalence classes $\ce_i$,
$i=1,\ldots ,n$ which do not communicate.

If $p_x >0 $ for all $x\in V$, then
from any $\eta\in \re_V$ we can reach the maximal configuration
$\eta^{max}$, therefore $\eta^{max}$ is recurrent and hence the Markov chain defined by
(\ref{2.9}) has only
one recurrent class containing the maximal configuration.

A subset $A$ of $\Omega_V$ is called {\sl closed under the Markov chain}
if for any $\eta\in A$ and $n\in\N$, $\pee_\eta (\eta_n\in A) =1$. A recurrent class is closed
under the Markov chain, and any set closed under the Markov
chain contains at least one recurrent class.
A probability measure $\mu$ on $\Omega_V$ is called invariant for the Markov
chain if for any $f:\Omega_V\rightarrow \R$ one has
\begin{equation}
\int (P_V f) d\mu = \int f d\mu.
\end{equation}
If the Markov chain has a unique recurrent class, then it also has a unique
invariant measure concentrating on that class and any initial probability measure converges
exponentially fast to this unique invariant measure.
In the next section we show that the invariant measure of
the Markov chain (\ref{2.9}) is the uniform probability measure on $\re_V$.

\section{The group of toppling operators}
In this section we show the group property of the addition operators working
on the set of recurrent configuration, and some related results on subsets of addition operators. For notational convenience we will skip the indices $V$ referring to the finite volume in what follows.

\medskip\noindent
By the abelian property, the set
\begin{equation}
\s = \{ \prod_{x\in V}a_x^{n_x}: n_x\in \N \}
\end{equation}
working on the set of all stable configurations is an abelian
semigroup.
We first show that $\s$ working on the set of recurrent configurations
is a group.
\begin{proposition}\label{sandgroup}
\begin{enumerate}
\item $\s$ restricted to $\re$ is an abelian group (denoted by $G$).
\item For all $x\in V$, there exist $n_x\geq 1$ such that for all $\eta\in \re$:
\begin{equation}
a_x^{n_x}\eta = \eta
\end{equation}
\item The cardinality of $G$ equals the cardinality of $\re$.
\item We have the following closure relation: for all $x\in V$
\begin{equation}\label{closure}
\prod_y a_y^{\Delta_{x,y}} = e,
\end{equation}
where $e$ denotes the neutral element in $G$.
\end{enumerate}

\end{proposition}
{\bf Proof:} First of all notice that $\eta\in \re$ and $g\in\s$ implies
(by positivity of the addition probabilities $p_x$) that $\eta\hookrightarrow g\eta$, and
hence $g\eta$ is recurrent. Therefore $\re$ is closed under the action of $\s$.
Let $\eta\in \re$. Since in the Markov chain (\ref{2.9})
we add on any site with positive probability, there exist $n_x \geq 1$ such that
\begin{equation}
\prod_{x\in V} a_x^{n_x (\eta)} \eta =\eta
\end{equation}
Consider the set
\begin{equation}
A=\{ \zeta\in\re:\prod_{x\in V} a_x^{n_x (\eta)}\zeta = \zeta \}
\end{equation}
This set is non-empty and by the abelian property, it is closed under the action of
the semigroup $\s$ and hence under the Markov chain. Therefore it contains
$\re$ and thus, by definition, equals $\re$. Hence the product
\begin{equation}
\prod_{x\in V} a_x^{n_x (\eta)}
\end{equation}
acts on $\re$ as the neutral element, and inverses of $a_x$ acting on
$\re$ are defined by
\begin{equation}
a_x^{-1} =  a_x^{n_x (\eta ) -1}\prod_{y\in V,y\not= x} a_y^{n_y (\eta)}
\end{equation}
This proves the group property. To prove statement (2) of the
proposition, note that $G$ is a finite group, so every element is of finite order. To prove
point (3), suppose that
$g\eta = g'\eta$ for some $\eta\in\re$, $g,g'\in G$. Then by abelianness:
\begin{equation}
g(h\eta) = g'(h\eta),
\end{equation}
for any $h\in G$. The set $\{ h\eta: h\in G\}$ is closed under the working of
$\s$, and contains $\eta$. Therefore it coincides with $\re$. We conclude that $g\zeta
=g'\zeta$ for any $\zeta\in\re$, and hence by definition of $G$ this implies $g=g'$. Therefore
the mapping
\begin{equation}
\Psi_\eta: G\rightarrow \re: g\mapsto g\eta
\end{equation}
is bijective.
Finally (as explained already in \cite{Dhar}) the closure relation is the
consequence of the observation that adding $\Delta_{x,x}$ grains to a site $x$ makes the site
topple, which results in a transfer of $-\Delta_{x,y}$ particles to any neighboring site $y$.
This gives
\begin{equation}
a_x^{\Delta_{x,x}}= \prod_{y\not= x} a_y^{-\Delta_{x,y}},
\end{equation}
which yields (\ref{closure}).
\QED
\begin{corollary}
The unique invariant measure of the Markov chain (\ref{2.9}) is the uniform measure on
$\re_V$.
\end{corollary}
{\bf Proof:} The invariant measure is unique since there is only one recurrent class. The
uniform measure is invariant under the working of any individual
addition operator $a_x$ because
\begin{equation}
\sum_{\eta\in\re_V} f(\eta) g(a_x \eta) = \sum_{\eta \in \re_V} f(a_x^{-1}\eta) g(\eta),
\end{equation}
and we can choose $f=1$. Hence the uniform measure on $\re$ is invariant under the working
of the Markov transition operator $P_V$ of (\ref{2.9}), independently
of the chosen $p$.\QED
{\bf Remark:} From the implication $\eta\in\re$, $g\in\s$, then $g\eta\in\re$,
it follows that $\eta\in\re$ and $\zeta \geq \eta$ implies $\zeta\in \re$.
\begin{definition}
Let $A\subset \Omega$ and $\s'\subset\s$. We say that $A$ has the
$\s'$-group property if $\s'$ restricted to $A$ forms a group.
\end{definition}
\begin{definition}
Let $\s'\subset\s$, and $A,B\subset\Omega$. We say that $A$ is $\s'$-connected
to $B$ if for any $\eta\in A$ there exists $g\in\s'$ such that
$g\eta\in B$.
\end{definition}
\begin{proposition}\label{gr}
Let $\s'\subset\s$, and $A\subset\Omega$. Suppose $A$ has the $\s'$-group
property and is $\s'$-connected to $\re$. Then $A$ is a subset of $\re$.
If, in addition, $A$ is closed under the action of $\s$, then $A$ equals $\re$.
\end{proposition}
{\bf Proof:}
Let $\eta\in A$. Then there exists $g\in\s'$ such that $\zeta=g\eta \in\re$.
Since $g$ acting on $A$ can be inverted, $\eta = g^{-1} \zeta$.
Therefore, $\zeta$ and $\eta$ communicate in the Markov chain.
Since $\zeta\in \re$, it follows $\eta\in \re$. Therefore, $A$ is
a subset of $\re$. If $A$ is closed under the action of $\s$, then
it is closed under the Markov chain, and hence contains $\re$.
\QED

\section{Recurrent configurations}\label{recsec}

We first show that Dhar's definition of recurrence in \cite{Dhar3} is the same as the
classical definition in terms of the Markov chain.
\begin{theorem}
We have the following identity:
\begin{equation}\label{bla}
\re = \{ \eta\in\Omega: \forall x\in V\  \exists n_x \geq 1:
a_x^{n_x}\eta = \eta \}
\end{equation}
\end{theorem}
{\bf Proof:}
Denote the set in the right hand site of (\ref{bla}) by $A$. Remark that the $n_x$ can be chosen
independent of $\eta$. Indeed, if
$a_x^{n_x (\eta)}\eta = \eta$ for all $\eta\in A$, then by abelianness,
for all $\zeta\in A$ we obtain
\begin{equation}
a_x^{\prod_{\eta\in A} n_x (\eta )} \zeta =\zeta.
\end{equation}
By Proposition \ref{sandgroup}, $\re\subset A$. Moreover, restricted
to $A$, inverses on $\s$ can be defined by $a_x^{-1}= a_x^{n_x-1}$. Therefore,
$\s$ restricted to $A$ is a group, and $A$ is clearly $\s$-connected to the
maximal configuration which belongs to $\re$. \QED

\noindent
The previous result showed that the recurrent configurations are precisely those, for which repeated
adding of grains at any vertex eventually  leads to the original configuration. The following  lemma is related. It shows that if we start with a configuration outside $\re$,
then by repeated addition at any particular vertex, we eventually obtain a recurrent configuration. We shall use this result later.
\begin{lemma}\label{onesitetop}
Define
\begin{equation}
\Omega'=\{ \eta\in\Omega: \forall x\in V\ \exists n_x: a_x^{n_x}\eta\in\re\}
\end{equation}
then $\Omega'=\Omega$.
\end{lemma}
{\bf Proof:} Certainly, $\Omega'$ is not
empty, since it contains $\re$. Define, for $x\in V$, the ``diminishing-operator" $\mbox{dim}_x(\eta)$ as follows:
\begin{equation}
\mbox{dim}_x (\eta)  (y) = \max \{(\eta (y) -\delta_{y,x}), 1\}.
\end{equation}
In words, we substract one from $\eta$ at site $x$, if this is possible.
We want to prove now that for $\eta\in\Omega'$, $\mbox{dim}_x (\eta)$ is still
in $\Omega'$. Since the maximal configuration $\eta^{\max}$ is in $\re$, this clearly implies the statement of the lemma. Let
$\eta\in\Omega'$.
Clearly
$a_x^{n_x+1} \mbox{dim}_x (\eta )= a_x^{n_x}\eta \in \re$. Now let $y\in V$. By adding at $y$
we can create as many topplings as we want at any site $z\in V$, i.e., we can write
\begin{equation}
a_y^k = \prod_{z\in V} a_z^{r_z (k)},
\end{equation}
where $r_z(k) \to \infty$ for any $z\in V$ as $k \to \infty$. Since $\eta\in\Omega'$, there exists $n_y $ such that $a_y^{n_y} \eta\in \re$. Now choose
$k>n_y$ big enough such that $r_x (k) \geq 1$, and $r_y (k) \geq n_y$.
Then we can write,
\begin{eqnarray*}
a_y^k \mbox{dim}_x (\eta ) &=& a_y^{n_y} \left(a_y^{r_y(k)-n_y}a_x^{r_x(k)-1}\prod_{z\in V, z \neq x,y} a_z^{r_z (k)} (a_x \mbox{dim}_x (\eta ))\right)\\
&= &a_y^{n_y} \left(a_y^{r_y(k)-n_y}a_x^{r_x(k)-1}\prod_{z\in V, z \neq x,y} a_z^{r_z (k)} (\eta)\right)\\
&=&\left(a_y^{r_y(k)-n_y}a_x^{r_x(k)-1}\prod_{z\in V, z \neq x,y} a_z^{r_z (k)}\right)a_y^{n_y}(\eta) \in \re.
\end{eqnarray*}
Hence we conclude that $\Omega'$ is closed under the $\mbox{dim}_x$-operation, for
any $x\in V$.\QED

\noindent
Next, we prove Dhar's formula for the number of recurrent
configurations (\cite{Dhar3}).
\begin{theorem}
\label{dharform}
$|\re| = \mbox{\rm det} (\Delta).$
\end{theorem}
{\bf Proof:}
Consider the following mapping:
\begin{equation}
\Psi: \Z^V\to G:n\mapsto \prod_x a_x^{n_x}.
\end{equation}
Clearly, $\Psi$ is a homomorphism, i.e., for $n,m\in\Z^V$,
\[
\Psi (n+m)= \Psi (n)\Psi (m).
\]
Since $\psi$ is also surjective, $G$ is isomorphic to the quotient $\Z^V/K$, where $K$ is the set
of those vectors $n\in \Z^V$ for which $\Psi (n)= e$.
By identity (\ref{closure}), we conclude
that
\begin{equation}
K\supset \Delta \Z^V,
\end{equation}
where
\begin{equation}
\Delta\Z^V = \{ \Delta n:n\in\Z^V\}
\end{equation}
Suppose now that $\Psi (n) =e$ for some $n\in \Z^V$. Then, writing $n=n^+-n^-$, where
$n^+ (x)\geq 0$, $n^- (x) \geq 0$ for all $x\in V$, we have
\begin{equation}\label{haaa}
\prod_x a_x^{n^+_x}=\prod_x a_x^{n^-_x}.
\end{equation}
Let $\eta\in\re$. By (\ref{haaa}), adding $n^+$ to $\eta$ gives the same result as adding
$n^-$. Therefore we can write
\begin{eqnarray}\label{lapla}
\eta + n^+ &=& \zeta + \Delta k^+\nonumber\\
\eta + n^- &=& \zeta + \Delta k^-,
\end{eqnarray}
where $k^+(x)$, resp $k^-(x)$ represents the number of topplings at site $x$ after addition
of $n^+$, resp.\ $n^-$. Subtracting the second from the first equation in (\ref{lapla}) leads
to the conclusion
\begin{equation}
n=n^+-n^-=\Delta (k^+-k^-),
\end{equation}
i.e., $K\subset \Delta\Z^V$.
We thus conclude that $G$ is isomorphic to $\Z^V/\Delta\Z^V$. The latter group has cardinality
$\mbox{det} (\Delta)$, as is well known. \QED

\begin{flushleft}
{\bf Remark:} 
\end{flushleft}
From the fact that each equivalence class of
$\Z^V/\Delta\Z^V$ can be identified with a unique recurrent
configuration, we deduce the following useful fact. If
$\eta\in\re$ is and we add to $\eta$ a configuration
$\zeta\in\N^V$ (point-wise addition) and 
$\xi\in\re$, $\alpha\in\N^V$ are such that
\begin{equation}\label{useful}
\eta+\zeta -\Delta\alpha = \xi,
\end{equation}
then this means the following: if we add to $\eta$ according to $\zeta$ , then we topple to $\xi$, and the number of topplings at each site is given by $\alpha$.

\section{Allowed configurations}

Let $\eta: V\rightarrow \N$ be a height configuration. For a subset $W\subset V$ we say that the restriction $\eta|W$ is a {\sl forbidden subconfiguration}
if for all $x$ in $W$ we have the inequality
\begin{equation}
\eta (x) \leq \mbox{deg}_W (x),
\end{equation}
where $\deg_W (x)$ denotes the number of neighbours of $x$ in $W$.
A configuration without forbidden subconfigurations is called {\sl allowed}.
The {\sl burning algorithm} determines whether a configuration
$\eta\in \Omega$ is allowed or not. It is described as follows: Pick
$\eta\in \Omega$ and erase all sites $x\in V$ satisfying the inequality
\[
\eta (x) > \sum_{y\in V,y\not= x} (-\Delta_{x,y}).
\]
This means ``erase the set $E_1$ of all sites $x\in V$ with a height
strictly larger than the number of neighbors of that site in $V$''.
Iterate this procedure for the new volume $V\setminus E_1$, and the new
matrix $\Delta^{V\setminus E_1}$ defined by
\begin{eqnarray*}
\Delta^{V\setminus E_1}_{x,y} &=& \Delta^V_{x,y} \ \mbox{if } x,y\in
V\setminus E_1\nonumber\\
&=& 0 \ \mbox{otherwise},
\end{eqnarray*}
and so on. If $\eta$ contains a forbidden subconfiguration, then the algorithm will never remove vertices
in this subconfiguration, and the limiting set is nonempty. On the other hand, if there is no such forbidden
subconfiguration in $\eta$, then the algorithm will
eventuallt remove all vertices. Hence in this case, the limiting set will be empty. So a configuration is allowed if and only of the burning algorithm erases (burns) all vertices. Let us denote by $\aaa$ the set of all allowed configurations.

\begin{lemma}
\label{jju}
\begin{enumerate}
\item The set of allowed configurations is closed under the action of $\s$.
\item $\aaa\supset \re$.
\end{enumerate}
\end{lemma}
{\bf Proof:} Let $\eta\in\aaa$. Addition on a site $x\in V$ for which $\eta(x) <\Delta_{x,x}$
increases the height and thus cannot create a forbidden
subconfiguration if the original $\eta$ does not contain a forbidden
subconfiguration. Suppose that by toppling the site $x$, we create a forbidden
subconfiguration in the subvolume $V_f\subset V$.
After toppling at site $x$, the new height at site $y$ satisfies
\begin{equation}
T_x\eta (y) = \eta (y) - \Delta_{x,y}.
\end{equation}
If $T_x\eta|V_f$ is a forbidden subconfiguration, then for all $y\in V_f\setminus \{x \}$ we
have
\[
\eta (y) \leq \mbox{deg}_{V_f} (y) + \Delta_{xy}
\]
i.e.,
 \[
\eta (y) \leq \mbox{deg}_{V_f\setminus \{x \}} (y)
\]
and we conclude that $\eta|V_f\setminus \{x \}$ is a forbidden subconfiguration for $\eta$,
which is not possible since $\eta$ was supposed
to be allowed. Since the operators $a_x$ are products of additions
and topplings, we conclude $\eta\in\aaa$ implies $a_x\eta\in\aaa$.
Clearly, the maximal configuration $\eta^{max}\in\aaa$. Therefore,
$g\eta^{max}\in\aaa$ for all $g\in\s$, and thus point (2) of the lemma
follows.\QED
The following lemma is called ``the multiplication by identity test"
(see e.g., \cite{Dhararch}
\begin{lemma}\label{test}
For $x\in V$, let $\alpha_x$ denote the number of neighbors of $x$
in $V$. The following two assertions are equivalent
\begin{enumerate}
\item
$\eta\in\aaa$
\item
$\prod_{x\in V} a_x^{\Delta_{x,x}-\alpha_x}\eta = \eta.$
\end{enumerate}
\end{lemma}
{\bf Proof:} Let $\eta\in\Omega$. Upon addition of $\sum_{x} (\Delta_{x,x}-\alpha_x)\delta_x$
to $\eta$, we have to topple those boundary sites
$x\in V$ that satisfy
the inequality
\begin{equation}
\Delta_{x,x} - \alpha_x + \eta (x) > \Delta_{x,x}.
\end{equation}
These are precisely the sites that can be burned in the ``first step"
of the burning algorithm. Let us call $B_1$ the set of those sites.
After toppling all sites in $B_1$, we will have a toppling at those
sites $x$ in $\partial V\setminus B_1$ that
satisfy the inequality
\begin{equation}\label{br}
\Delta_{x,x}- \alpha_x + \eta (x) + \alpha_x^{B_1} > \Delta_{x,x}
\end{equation}
where $\alpha_x^{B_1}$ denotes the number of neighbors of $x$ in $B_1$.
(\ref{br}) is equivalent to
\begin{equation}
\eta (x) > \alpha_x^{V\setminus B_1}.
\end{equation}
Those sites that topple after the toppling of sites in $B_1$ thus
coincide with the sites that can be burned after burning of $B_1$.
Continuing this reasoning, we arrive at the conclusion that $\eta$
does not contain a forbidden subconfiguration if and only if upon addition
of $\sum_x (\Delta_{x,x}-\alpha_x)\delta_x$ every site topples {\sl at least}
once. We now show that for {\sl any} configuration, any site topples
also {\sl at most once} upon addition of $\sum_x (\Delta_{x,x}-\alpha_x)\delta_x$.
By the abelian property, it suffices to show this for the maximal
configuration. Since the maximal configuration is recurrent, it is sufficient
to prove the following equality (see (\ref{useful})):
\begin{equation}
\eta^{max} (x) + \left(\sum_{y} (\Delta_{y,y} -\alpha_y) \delta_y\right) (x)
-\sum_{y} \Delta_{y,x} = \eta^{max} (x),
\end{equation}
or
\begin{equation}
\sum_{y} \Delta_{y,x} = \Delta_{x,x} - \alpha_x,
\end{equation}
which is obvious. Therefore we conclude $\eta\in\aaa$ to be equivalent with the fact that upon
addition of $\sum_x (\Delta_{x,x}-\alpha_x)\delta_x$,
every site topples {\sl precisely} one time, and hence the resulting configuration is $\eta$.
\QED
\begin{corollary}\label{partial}
Consider the following subset of $\s$:
\begin{equation}
\s_\partial = \{ \prod_{x\in\partial V} a_x^{n_x}: n_x\in \N \}.
\end{equation}
Restricted to $\aaa$ , $\s_\partial$ defines an abelian group.
\end{corollary}
{\bf Proof:}
By Lemma \ref{test}, restricted to $\aaa$, $\s_\partial$ has the neutral
element
\begin{equation}\label{*}
\prod_{x\in\partial V} a_x^{\Delta_{x,x}-\alpha_x} =e.
\end{equation}
Because in the product (\ref{*}) every operator appears with a power at least
one, inverses of the boundary operators are defined by (\ref{*}) and abelianness.\QED

\noindent
Finally, we can now prove the fact that ``allowed" is the same as ``recurrent"
\begin{theorem}\label{all=rec}
$\aaa = \re$
\end{theorem}
{\bf Proof: } By Corollary \ref{partial}, $\s_\partial$ restricted to
$\aaa$ is a group. By Lemma \ref{onesitetop}, $\aaa$ is $\s_\partial$-connected to $\re$. Therefore, the theorem follows as an application of Proposition \ref{gr}.\QED
{\bf Remark:} From combination of Proposition \ref{gr} and Lemma \ref{onesitetop}, we obtain the following generalization of the previous theorem. If $A$ is any set closed under the action of $\s$, and has
the $\s'$-group property for some $\s'\subset\s$, then $A=\re$.

\end{document}